\documentclass[twocolumn,twoside,slac_two]{revtex4}
\usepackage{graphicx}
\usepackage{fancyhdr}
\begin{document}

\title{Prospects for the first Top pair cross section measurement in the semileptonic and dilepton channels at CMS}

%

\author{Ashish Kumar (for the CMS Collaboration)}
\affiliation{Department of Physics, State University of New York at Buffalo, Buffalo, NY 14260, USA}

\begin{abstract}
Although the top quark has been discovered in 1995 and studied extensively by the Tevatron experiments, the top quark
will remain special for years to come due to unique opportunities it offers. Because of the large top-antitop production 
cross section and high luminosity, the LHC would be a Top factory, 
producing a large sample of top quarks even at the initial low luminosities. This will enable a rich program of top quark 
physics to be explored, both within the Standard Model and using top quarks as probes of physics beyond 
the Standard Model. Prospects for the observation of top pair production in the proton-proton collisions at the
center of mass energy $\sqrt{s}$=10 TeV in the dilepton and lepton+jets final state are discussed. The
emphasis is put on analysis strategies for the early phase of CMS operation with data corresponding
to integrated luminosities of 10-20 pb$^{-1}$ considering a realistic detector performance. 
\end{abstract}

\maketitle

\thispagestyle{fancy}


\section{Introduction}
In the Standard Model (SM), the top quark has a special place among the constituents of matter because of its 
unique properties: it has much larger mass with respect to all the other fermions, its decays involve real 
(rather than virtual) $W$ bosons, and it decays long before it can hadronize thereby allowing to preserve information about its 
spin and polarization state. The large value of the top mass, which is close to the 
electroweak symmetry breaking scale, may point to a special role that top quark plays in the 
symmetry breaking. Since the discovery of the top quark in 1995, all our current knowledge about its 
production and properties (such as its mass, width, spin, charge and couplings to other particles, etc.) 
come from the CDF and D\O~experiments at the Fermilab Tevatron collider. However, besides the precise 
measurement of the mass, $m_{top} = 173.1\pm1.3$ GeV~\cite{tevew}, with a relative uncertainty of 0.75\%, and the measurement 
of the top pair production cross section, $\sigma (pp\rightarrow t\bar{t})=7.50\pm0.48$ pb for $m_{top}$ = 172.5 GeV~\cite{cdfpublic}, 
with a relative uncertainty of 7\% that reaches the current accuracy of the NLO QCD calculations, the rest of 
measurements of top quark properties are still very statistically limited~\cite{cdfpublic,d0public}.  
At the Large Hadron Collider (LHC) top quarks will be produced copiously. With seven times larger center of 
mass energy at $\sqrt{s}=$14 TeV and higher luminosity, more than 8 million top pairs and 2 millions of single top 
events will be produced per year of nominal data-taking (integrated luminosity of 10 fb$^{-1}$). Consequently, the LHC experiments 
will herald a new era of precision measurement in the top quark sector.  
This wealth of statistics will allow the detailed investigation of its production and properties, 
providing both stringent tests of the SM and opportunities for new physics searches~\cite{cmstdr}.\\

The measurement of the top quark pair production cross section will be one of the early LHC physics goals 
(even without b-tagging) as it would provide test of the theoretical predictions in the new energy regime. 
The large top quark sample available from the start of LHC will also play an important role in commissioning 
the detectors during the first data-taking period. Given the well established decay properties and their 
final state topologies, $t\bar{t}$ events will also constitute one of the main benchmark samples in many fields, 
from jet energy scale determination to the measurements of the performance for b tagging and lepton 
identification tools. The higher multiplicities and transverse momenta of jets in $t\bar{t}$ events, compared to 
other SM processes, make the calibration environment more similar to the one expected in many New Physics searches. 
Furthermore, the top signal being one of the most important sources of background to other new physics signal, 
a detailed understanding of the top production rates and decay properties will be a necessary path to new discoveries.

\section{Top Production at LHC}
\begin{figure}[h]
\centering
\includegraphics[width=80mm]{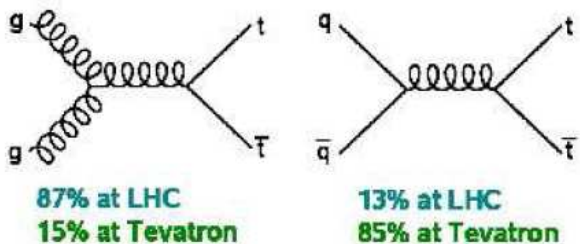}
\caption{The Feynman diagrams of the leading order processes for Top pair production at the Tevatron and LHC.} \label{diagrams}
\end{figure}

At the LHC, top quarks are dominantly produced in $t\bar{t}$ pairs via the processes  $q\bar{q} \rightarrow t\bar{t}$
and $gg \rightarrow t\bar{t}$ (Figure~\ref{diagrams}). Due to the larger centre-of mass energy available at the LHC, 
typical momentum fractions of the incoming partons are very low with $x\sim$0.025 where the gluon density 
of the proton dominates. Hence, about 87\% of the $t\bar{t}$ production comes from the gluon-gluon fusion process
and the remaining 13\% comes from quark-antiquark annihilation, which is dominant at the Tevatron. 
In contrast to near threshold production of $t\bar{t}$ at the Tevatron, at the LHC, the relative difference in the 
$x$ values of the colliding partons can be quite large, resulting in a strong forward boost of 
the $t\bar{t}$ system. At $\sqrt{s}$=14 TeV, the expected top pair cross-section at the next-to-leading order (NLO) is 
$\sigma(t\bar{t})= 908\pm$ 88 pb for $m_{top}$ = 175 GeV~\cite{lhctop2}. Compared to the Tevatron, the production cross section 
at the LHC is expected to be two orders of magnitude 
larger for top quark pairs, but only an order of magnitude larger for background events such as $W$+jets events, 
giving large improvements in both the available statistics and the signal to background ratio.\\

In the SM, once produced, a top quark decays for nearly 100\% into a $b$-quark and a $W$-boson. The $W$-boson then 
decays either into a pair of light, i.e. non-$b$, quarks ($\sim$2/3 of the time) or into a lepton and neutrino ($\sim$1/3 of the time). 
The $t\bar{t}$ final states result from the subsequent decay modes of the two $W$ bosons. 
The resulting channels are therefore called fully hadronic, semi-leptonic and fully 
leptonic with branching ratios of 46\%, 44\%, and 10\%, respectively. \\

The LHC operations was scheduled to restart in mid-November this year with a clear goal for a long 
first physics run through winter 2009 to autumn 2010 at $\sqrt{s}=$10 TeV. According to the latest 
schedule, the LHC will initially run at 3.5 TeV per beam to gain experience of running the machine safely 
and then smoothly ramp up towards 5 TeV per beam. The estimated $t\bar{t}$ cross section is
$\sigma(t\bar{t})=414 \pm$ 41 pb at $\sqrt{s}=$10 TeV~\cite{lhctop2}. 
Although the top pair cross section is reduced by about a half at this energy, nevertheless the
top physics program will remain quite competitive.
Extensive efforts have been put to commission the CMS detector by taking data with cosmic triggers. 
These data allowed to gain experience in issues related to detector timing, alignment and calibration and therefore, 
to improve its readiness towards recording the first collision data. 

\section{Establishing the Top signal with early CMS data}
The CMS experiment~\cite{cms} has developed simple and robust analyses to establish the top signal with the very 
early dataset corresponding to lowest initially available integrated luminosities of about 10-20 pb$^{-1}$ at $\sqrt{s}$=10 TeV.
The focus has been on channels with leptonic $W$ decay(s) without using tools such as $b$-tagging and even 
missing transverse energy which might not be reliable in the early running phase when these aspects of 
the detector performance are not well understood.

\subsection{Di-lepton channel: $t\bar{t}\rightarrow W(\rightarrow e/\mu+\nu_{e/\mu})bW(\rightarrow e/\mu+\nu_{e/\mu})b$}
The dileptonic final state is a rare signature where both $W$ bosons, produced by the decay of the top pair, 
decay leptonically to an electron or muon giving rise to two high-transverse momentum ($p_T$ ) leptons, 
two energetic jets from the hadronization of $b$ quarks and large missing transverse energy $E_T^{miss}$ (due to large momentum 
imbalance in the plane transverse to the beam) from the two undetected neutrinos ($\nu$). Although the branching ratios
for the dileptonic channels is small ($\sim$5\%), this is possibly the first place where 
the evidence of top events can be seen because of the cleaner signal. The processes expected to contribute to the background 
to this experimental signature are Drell-Yan+jets ($DY+$jets), dibosons, $W$-bosons with jets, single-top quark, and QCD multi-jet.
The main background comes from DY+jets production. The lepton fake contribution from QCD multi-jets is small thanks to 
the requirement of the presence of two leptons in the final state. In events with an electron and a muon in the final state, 
even the $DY$ events will not contribute to the background. The CMS analysis presents~\cite{dilep} the strategies designed to provide 
the first measurement of  the cross section in the dilepton final state with a data sample of 10 pb$^{-1}$. It relies
on simple counting experiment approach and the excess of event candidates passing a selection of characteristic signal over the 
expected contribution from the background processes is ascribed to the production of $t\bar{t}$. It also applies 
data-driven methods to estimate the background contributions which can not be reliably controlled using simulation. 
For example, $DY \rightarrow e^+e^-$ and $\mu^+\mu^-$ events can mimic the signal due to large mismeasurement of the $E_T^{miss}$ or of the
lepton momenta. Also, due to misidentification of jets into isolated lepton (fake lepton), QCD multi-jet events can resemble the
$t\bar{t}$ signature.\\

\begin{figure*}[t]
\centering
\includegraphics[width=135mm]{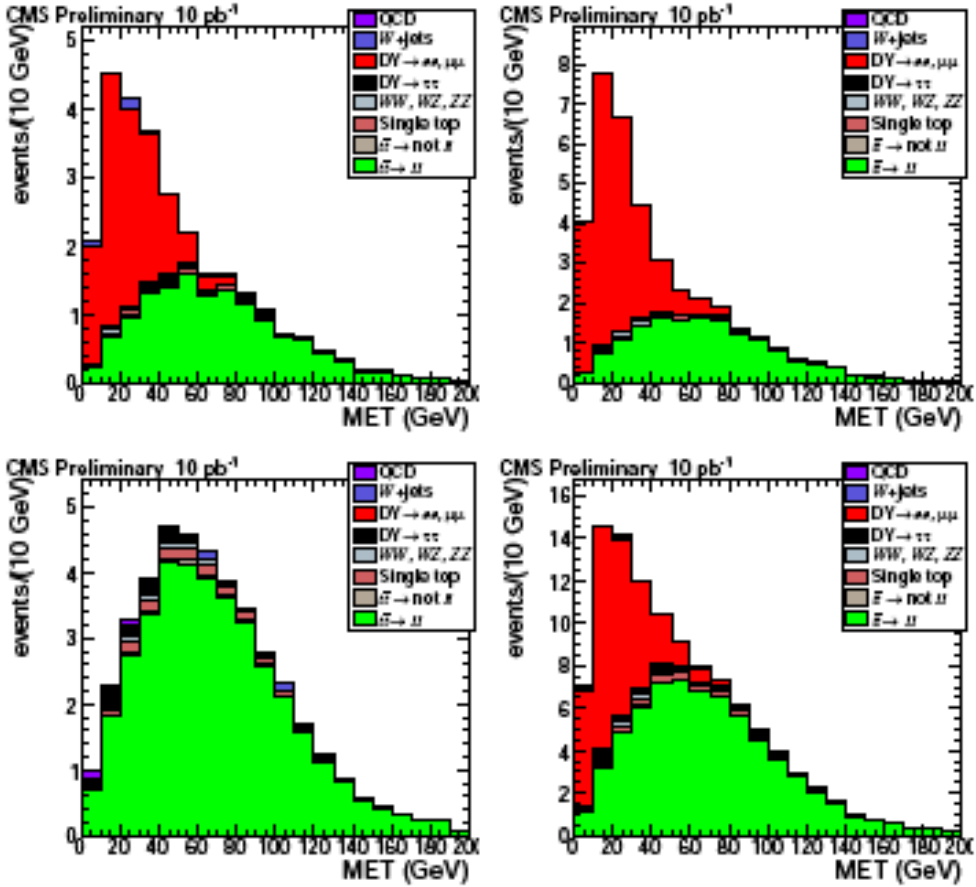}
\caption{The expected number of dilepton events as a function of $E_T^{miss}$ normalized to 10 pb$^{-1}$ in $e^+e^-$ (top-left), $\mu^+\mu^-$ (top-right), $e\mu$ (bottom-left), and all channels combined (bottom-right). The distributions are for events passing dilepton selections and having at least two jets.} \label{met_ll}
\end{figure*}

\begin{figure*}[t]
\centering
\includegraphics[width=135mm]{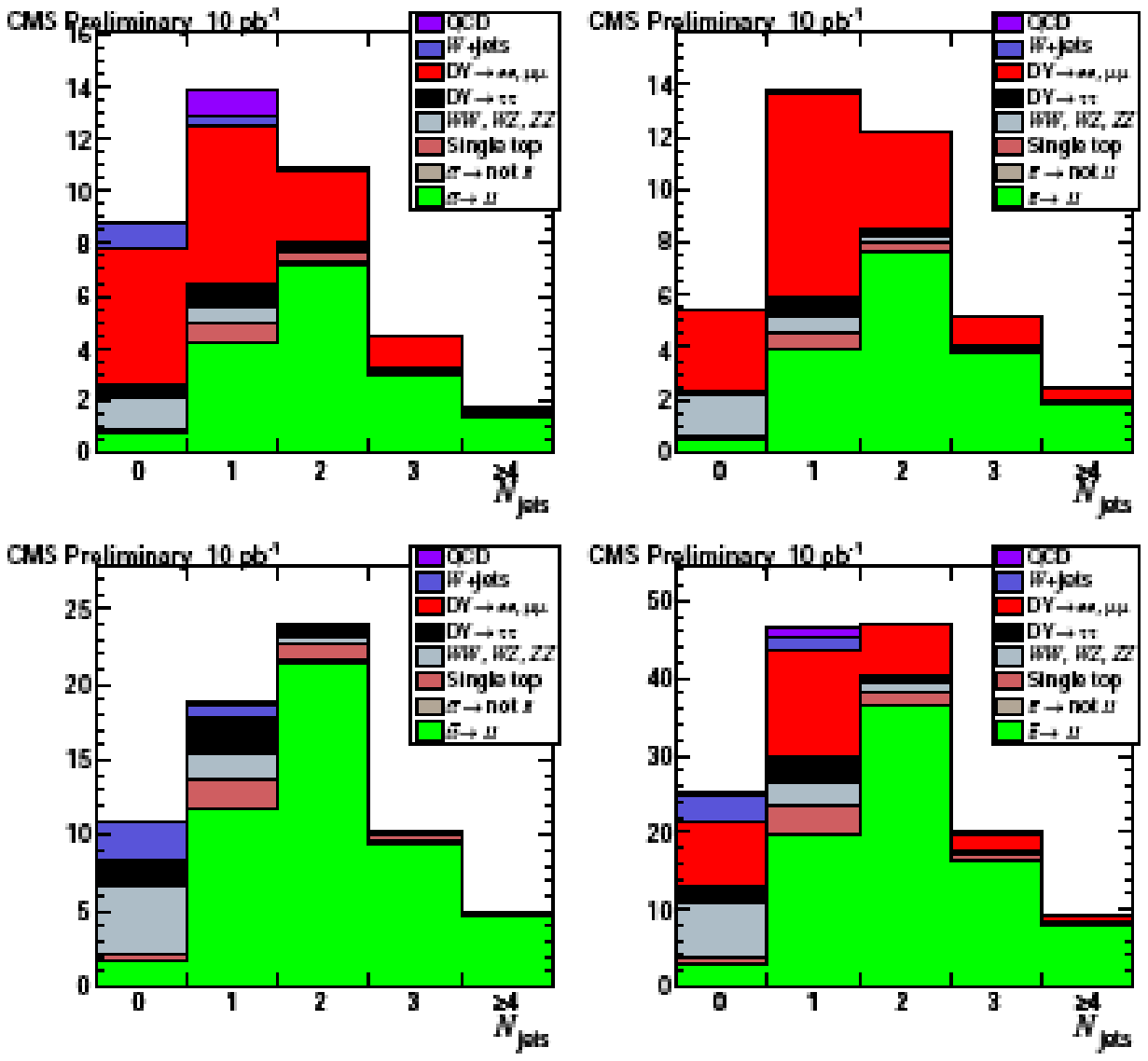}
\caption{The expected number of dilepton events as a function of jet multiplicity normalized to 10 pb$^{-1}$ in $e^+e^-$ (top-left), $\mu^+\mu^-$ (top-right), $e\mu$ (bottom-left), and all channels combined (bottom-right).} \label{njet_ll}
\end{figure*}

The events are required to pass single electron or muon trigger with  $p_T$ threshold of 15 GeV and 
9 GeV, respectively. The offline lepton selection required two opposite-sign charge leptons with isolation criteria based 
on the calorimeter and the tracking satisfying $p_T>$20 GeV and $|\eta|<$2.4. Three exclusive dilepton final states are 
considered : two electrons ($e^+e^-$), two muons ($\mu^+\mu^-$), and an electron and a muon ($e^{\pm}\mu^{\mp}$).
At least two jets are  required in the event which are reconstructed in the calorimeter using a seedless 
infrared-safe cone jet algorithm  with a cone size of $\Delta R$=0.5. 
Events with an $e^+e^-$ or $\mu^+\mu^-$ pair with invariant mass within 15 GeV of the $Z$ mass are rejected.
This $Z$-veto reduces significantly the $DY$+jets background. After the $Z$-veto, the ability to further reject the 
$DY$+jets background depends crucially on the performance of $E_T^{miss}$ measurement.
The $E_T^{miss}$ requirement in $e^+e^-$ or $\mu^+\mu^-$ channels is $>$30 GeV and $>$20 GeV in the $e\mu$ channel.
The distribution of the $E_T^{miss}$ in events with at least two jets is shown in Fig.~\ref{met_ll} where the $t\bar{t}$ events
with significantly large $E_T^{miss}$ can be clearly seen. Fig.~\ref{njet_ll} shows the expected number of events passing the 
main event selections normalized to 10 pb$^{-1}$ where the $t\bar{t}$ signal is clearly visible for $N_{jets}\geq$2.
The signal purity in $e\mu$ events is outstanding. Clear observation of the signal is expected in the 10 pb$^{-1}$
sample with a signal-to-background ratio ($S/B$) of 4 to 1 in all channels combined and about 9 to 1 in the $e\mu$ channel alone.
The signal production cross section is expected to be measured with a  statistical uncertainty of 15\% and a systematic uncertainty
close to 10\% excluding the uncertainty on the integrated luminosity which is expected to be 10\%. 
The $S/B$ in $e^+e^-$ and $\mu^+\mu^-$ channels is about 2 to 1 where the dominant background originates from $DY+$jets, 
and the estimation of this background will depend on understanding of the $E_T^{miss}$ as well as the jet multiplicity. 
If this background cannot be controlled with the first data, the analysis can be limited to the much 
cleaner $e\mu$ channel, in which case the statistical uncertainty is expected to increase from 15\% to 18\%.

\subsection{The semileptonic channel: $t\bar{t} \rightarrow W(\rightarrow e/\mu+\nu_{e/\mu})bW(\rightarrow qq^{'})b$}
The 30\% of events where one W-boson decays hadronically (2 jets) while the other decays leptonically (muon/electron 
and a neutrino) is considered the golden channel as it has a very characteristic experimental 
signature that allows to obtain a clean sample of top events. Semi-leptonic top events have an (isolated) high-$p_T$ 
electron or muon, large $E_T^{miss}$, four high-$p_T$ jets, of which two jets originate originate 
from b-quark fragmentation. The major backgrounds to this channel can be broadly divided 
into two kinds: (i) backgrounds with a real prompt electron or muon, e.g. W+jets (where the W boson decays to electron/muon 
and a neutrino), Z+jets (where the Z boson decays to an electron/muon pair) and single top production 
(where the top quark decays semileptonically to electron/muon); (ii) backgrounds with fake or secondary 
electrons/muons arising from QCD multi-jet events. The CMS analyses~\cite{ejets,mujets} present the strategies designed to provide 
the first measurement of the cross section in the lepton+jets final state with a data sample of 20 pb$^{-1}$.\\

\begin{figure}[h]
\centering
\includegraphics[width=80mm]{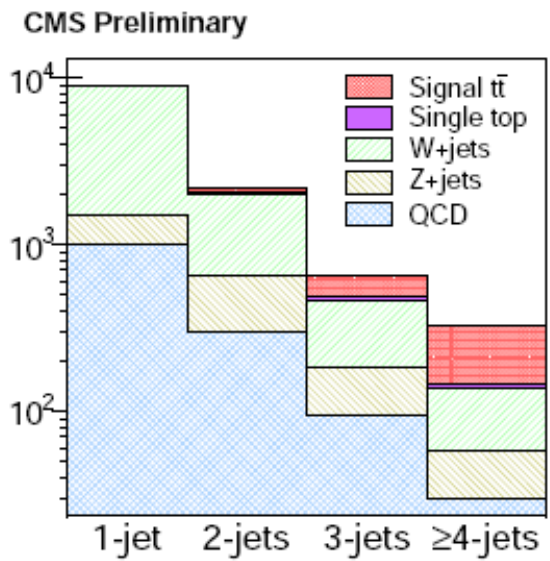}
\caption{Expected number of signal and background events in the $e+$jets channel as a function of jet multiplicity normalized to 20 pb$^{-1}$.} \label{njet_ejets}
\end{figure}

\begin{figure}[h]
\centering
\includegraphics[width=80mm]{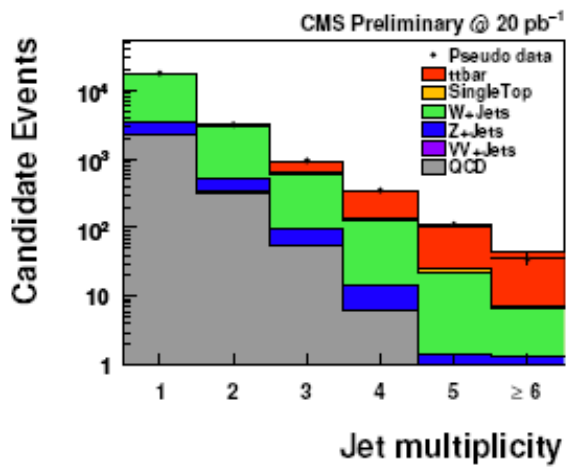}
\caption{Expected number of signal and background events in the $\mu+$jets channel as a function of jet multiplicity normalized to 20 pb$^{-1}$.} \label{njet_mujets}
\end{figure}

The event selection starts with the requirement of an inclusive single electron (muon) trigger with a high $E_T$ 
threshold of 15 (9) GeV. The subsequent offline selection requires a $t\bar{t}$ candidate event to have one 
reconstructed electron (muon) with $p_T>$30 (20) GeV, $|\eta|<$ 2.5 (2.1) and at least four jets 
with $p_T >$ 20 GeV and $|\eta|<$ 2.4. The leptons are required to be isolated by making use of calorimeter and tracker 
based isolation variables. Events with any additional electrons or muons are vetoed in order to reduce the 
contamination from dileptonic top decays, which are treated as background here, as well as from $Z+$jets and diboson events. 
The electron+jets and muon+jets channel are made statistically independent from each other, and so events 
with a good, isolated electron with $p_T >$ 30 GeV are rejected in the muon channel and vice versa in the electron channel. 
Even after the above selection, electron channel faces a significant contribution from QCD multi-jet events.
In order to further reduce this background, electrons are required to be within the barrel region of $|\eta|<$1.442. 
Since most of the material before the calorimeters in CMS is in the forward region, omitting the endcaps reduces 
the number of electrons from conversions considerably.\\

\begin{figure}[h]
\centering
\includegraphics[width=80mm]{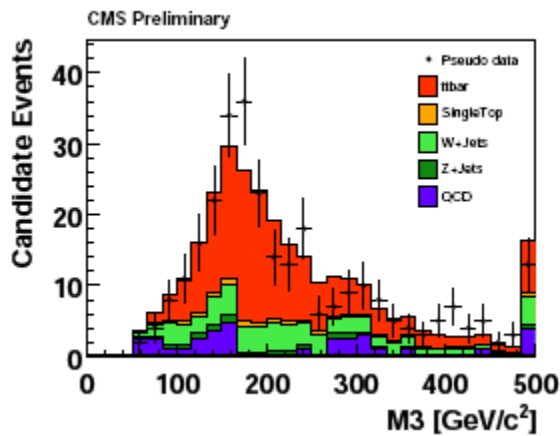}
\caption{Invariant mass of the three-jet combination with the highest vector sum $p_T$ for the events passing all selection in the $e+$jets channel.} \label{m3_ejets}
\end{figure}

\begin{figure}[h]
\centering
\includegraphics[width=80mm]{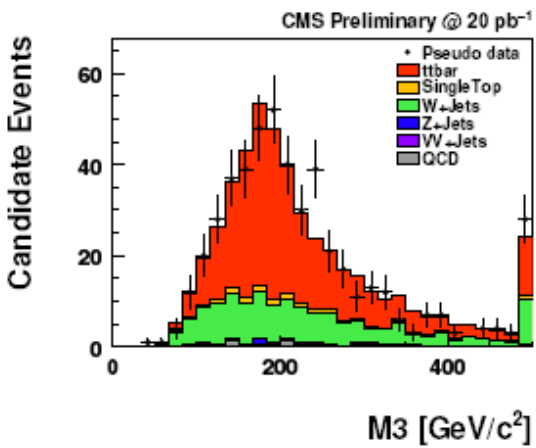}
\caption{Invariant mass of the three-jet combination with the highest vector sum $p_T$ for the events passing all selection in the $\mu+$jets channel.} \label{m3_mujets}
\end{figure}

In the electron channel~\cite{ejets}, the final selection yields 172 $t\bar{t}$ candidate events, with the background event yield of 
108 events leading to a $S/B$ of 1.6. Among the background events, $W+$ jets provides the major contribution with 57 events. 
In the muon channel~\cite{mujets}, the final selection yields 320 $t\bar{t}$ candidate events, with the background 
event yield of 171 events leading to a $S/B$ of 1.9. As expected, $W+$ jets provides the main background with 140 events. 
An estimation of the QCD multi-jet  contribution in the final selection was performed in a data-driven way using an extrapolation 
method with the lepton isolation distribution. Figure~\ref{njet_ejets} and Fig.~\ref{njet_mujets} show the expected 
number of signal and background events in the electron and muon channel as a function of jet multiplicity.
For jet multiplicity of four and higher, the sample is dominated by $t\bar{t}$ signal events, while the lower jet bins
are dominated by background events.\\

In the absence of $b$-tagging, the analyses rely on 
kinematic information to extract the top signal. 
The invariant mass of hadronically decaying top quark candidates ($M3$), which is formed by selecting 3-jet 
combinations with the highest vector sum of the jet $p_T'$'s, is shown in Fig.~\ref{m3_ejets} and Fig.~\ref{m3_mujets} 
for the selected events in the electron and muon channel.  
It has a clear peak near the top quark mass and the discriminating power between signal and the W/Z+jets 
background can be seen.  The signal contribution is estimated by performing four parameter binned likelihood 
fit to data to extract $N_{t\bar{t}}$, $N_{W/Z+jets}$, $N_{single top}$, and $N_{QCD}$ using the $M3$ templates 
derived from Monte Carlo simulation. Since the shape of $Z+$jets is similar to $W+$jets, the $W+$jets template is used 
to fit both $W+$jets and $Z+$jets events. The QCD template is obtained from data in the non-isolated electron control region. 
With a 20 pb$^{-1}$ data, it is estimated that the $t\bar{t}$ cross section can be measured in the electron (muon) channel 
with 23\% (12 to 18\%) statistical and around 20\% (20 to 25\%) systematic error, dominated by the jet energy scale uncertainty.
In the  $\mu$+jets channel, a multivariate analysis technique based on boosted decision trees has also been used
which uses the kinematic and topological information of the event to distinguish the $t\bar{t}$ signal from
background processes~\cite{mujets-mva}. With this approach, the expected statistical and systematic uncertainties on the 
cross section measurement is $\sim$9\% and $\sim$22\% respectively excluding the luminosity uncertainty.

\section{Conclusion}
Top quark physics will play an important role in detector commissioning in the early days of the data taking 
of the CMS experiment at the LHC. Clear observation of the top signal is expected in the dilepton channel using 
simple counting experiment with 10 pb$^{-1}$ of data at $\sqrt{s}$=10 TeV. The $t\bar{t}$ cross section is expected
to be measured with statistical uncertainty of 15\% and a systematic uncertainty
close to 10\% excluding the uncertainty on the integrated luminosity which is expected to be 10\%.
Top signal could also be established in lepton+jets 
channel with 20 pb$^{-1}$ of data at $\sqrt{s}$=10 TeV where the top is reconstructed from the 3-jet combination with highest 
vector sum $p_T$. The $t\bar{t}$ cross section can be measured in the electron (muon) channel 
with 23\% (12 to 18\%) statistical and around 20\% (20 to 25\%) systematic error, dominated by the jet energy scale uncertainty.
In the $\mu$+jets channel, application of a multivariate analysis technique which employs the kinematic 
and topological information of the event to distinguish the $t\bar{t}$ signal yields cross-section measuement with the 
statistical and systematic uncertainties of $\sim$9\% and $\sim$22\% respectively. 
Understanding of the systematic effects and assessment of background expectations will improve after data collection begins.
With increasing integrated luminosities, the $b$-quark content of the events will be exploited to obtain 
better signal to background ratio.

%


\begin{acknowledgments}
I would like to thank Tim Christiansen and Claudio Campagnari for the help with resources during the
preparation of the talk. Additional useful discussions with Avto Kharchilava, Boaz Klima and Meenakshi Narain
are gratefully acknowledged.
\end{acknowledgments}

\bigskip 

\begin{thebibliography}{9}   

\bibitem{tevew}   Tevatron Electroweak Working Group, arXiv:hep-ex/09032503
\bibitem{cdfpublic} http://www-cdf.fnal.gov/physics/new/top/top.html
\bibitem{d0public} http://www-d0.fnal.gov/Run2Physics/top/
\bibitem{cmstdr}  The CMS Collaboration, CERN/LHCC 2006-001/021 CMS TDR 8.1/8.2.
\bibitem{lhctop2} M. Cacciari, S. Frixione, M. L. Mangano, P. Nason, and G. Ridolfi, JHEP 0809, 127 (2008).
\bibitem{cms} The CMS Collaboration, S Chatrchyan et. al, JINST 3 S08004 (2008).
\bibitem{dilep}  The CMS Collaboration, CMS Physics Analysis Summary TOP-09-002.
\bibitem{ejets}  The CMS Collaboration, CMS Physics Analysis Summary TOP-09-003.
\bibitem{mujets} The CMS Collaboration, CMS Physics Analysis Summary TOP-09-004.
\bibitem{mujets-mva} The CMS Collaboration, CMS Physics Analysis Summary TOP-09-010.

\end{thebibliography}

\end{document}